\documentclass[aps, pra, twocolumn, superscriptaddress, amsmath, tightenlines, longbibliography]{revtex4-1}

\usepackage{amssymb}
\usepackage{amsmath}
\usepackage{dcolumn}
\usepackage{graphicx}
\usepackage{mathrsfs}
\usepackage{appendix}
\usepackage{graphicx}
\usepackage{booktabs}
\usepackage{colortbl}

\definecolor{Dred}{RGB}{190,0,0}

\usepackage{url}
\usepackage[colorlinks]{hyperref}
\hypersetup{%
	plainpages=true,
	breaklinks=true,       
	hypertexnames=false,  
	pageanchor=true,
	colorlinks=true,
	linkcolor={blue},
	citecolor={red},
	urlcolor={blue},
	anchorcolor={black}
}

\def \hide#1{}

\begin{document}
\title{Unconventional Quantum Electrodynamics with Hofstadter-Ladder Waveguide}


\author{Xin Wang}
\affiliation{Institute of Theoretical Physics, School of Physics, Xi'an Jiaotong University, Xi'an 710049, People’s Republic of China}

\author{Zhao-Min Gao}
\affiliation{Institute of Theoretical Physics, School of Physics, Xi'an Jiaotong University, Xi'an 710049, People’s Republic of China}

\author{Jia-Qi Li}
\affiliation{Institute of Theoretical Physics, School of Physics, Xi'an Jiaotong University, Xi'an 710049, People’s Republic of China}

\author{Huai-Bing Zhu}
\affiliation{Institute of Theoretical Physics, School of Physics, Xi'an Jiaotong University, Xi'an 710049, People’s Republic of China}

\author{Hong-Rong Li}
\affiliation{Institute of Theoretical Physics, School of Physics, Xi'an Jiaotong University, Xi'an 710049, People’s Republic of China}

\date{\today}

\begin{abstract}
We propose a novel quantum electrodynamics (QED) platform where quantum 
emitters interact with a Hofstadter-ladder waveguide. We demonstrate several 
intriguing phenomena stemming from the nontrivial dispersion relation and 
vacuum mode properties led by the effective spin-orbit coupling. First, by 
assuming emitter's frequency to be resonant with the lower band, we find that 
the spontaneous emission is chiral with most photonic field decaying 
unidirectionally. Both numerical and analytical results indicate that the 
Hofstadter-ladder waveguide can be engineered as a well-performed chiral 
quantum bus. Second, the dynamics of emitters of giant atom form is explored by 
considering their frequencies below the lower band. Due to quantum 
interference, we find that both the emitter-waveguide interaction and the 
amplitudes of bound states are periodically modulated by giant emitter's size. 
The periodical length depends on the positions of energy minima points induced 
by the spin-orbit coupling. Last, we consider the interaction between two giant 
emitters mediated by bound states, and find that their dipole-dipole 
interaction vanishes (is enhanced) when maximum destructive (constructive) 
interference happens.  
\end{abstract}
\maketitle

\section{introduction}
The interactions between quantum emitters and the unavoidable baths with large degrees of freedom are the central topic of quantum optics~\cite{cohen1998atom,Clerk10}. For example, in the present of a vacuum bath, the emitter will spontaneously decay to its ground state as well as its frequency being renormalized due to Lamb shifts~\cite{Lamb1947,Scully1997}. By shaping the size of the environment or narrowing its spectrum bandwidth, many intriguing phenomena, such as isotropic propagation of photons and non-Markovian evolution arise~\cite{Quang1994,Lambropoulos2000,Giraldi2011,Lodahl2015,Stewart2020,Ferreira2021}. As discussed in Refs.~\cite{Petersen2014,Bliokh2015a,Bliokh2015b,Lodahl2017,Lang2022}, chiral emission can be observed via the subwavelength confinement in nanophotonic systems, which opens the possibilities to realize cascaded quantum networks. Moreover, when considering an emitter coupling to the bandgaps of a bath~\cite{John1990,Goban2014,GonzlezTudela2015,Douglas2016,Liu2017,Chang2018,Wang2021}, photonic bound states (in the form of an evanescent field) emerge~\cite{Douglas2015}. In this scenario, between atoms there are long-range dipole-dipole interactions by exchanging the virtual photons in the waveguide~\cite{Shahmoon2013,Ying2019}. 

In recent years, exploring quantum electrodynamics (QED) with emitters coupling to structured lattice environments, has attracted a lot of interests~\cite{Ramos2016,Calajo2016,Tudela2018,Tudela2019L,Leonforte2021}. Those artificial lattice reservoirs are widely studied in condensed matter physics, and usually have unconventional spectra, or topological properties with nontrivial vacuum modes.
In Refs.~\cite{GonzlezTudela2017,Tudela2017L}, by considering emitters 
interacting with a 2D tight-binding lattice environment, the authors showed 
that both superradiance and subradiance of collective atoms emerge in the 
nonperturbative regime. The unusual chiral bound states and directional 
dipole-dipole interaction were also demonstrated in a topological waveguide QED 
system~\cite{Bello2019,Kim2021}. In 
	Refs.~\cite{Roccati22,2022Bound}, the authors discussed how to realize 
	bound 
	states and 
	dipole-dipole interactions in  
	non-Hermitian photonic lattices.
All these studies indicate those structured lattice reservoirs with reduced dimensionality are versatile toolboxes for exploring novel quantum optical phenomena, as well as the potential applications in quantum information processing.

In artificial baths with the spin-orbit coupling, the 
motion and spin freedoms of a particle are linked, and many anomalous phenomena such as spin-Hall effect and topological insulators can be observed~\cite{Zhang2000,Murakami2003,Sinova2004,Wunderlich2005,Kane2005,Galitski2013,Zhou2013,Wu2016,Kartashov2017,Livi2017,Liu2011}. Since neither the spin nor the momentum is the well-defined quantum number to describe the dispersion relation, the spin-orbit coupling will produce nontrivial energy bands and photonic modes~\cite{Sala2015,Salerno2017}. 
The quantum optics with emitters interacting with baths of the spin-orbit 
coupling, is rarely studied. It is a simple but interesting toy
model in condensed matter physics (see 
Fig.~\ref{fig1m})~\cite{Creutz2001,Narozhny2005,Jaefari2012,Atala2014,Tai2017,Yuan2019,Guan2020}.
 As discussed in Ref.~\cite{Hugel2014}, the ladder contains two legs which play 
the roles of two freedoms in an effective spin. In the present of synthetic 
gauge fields, the effective spin will be locked to momentum freedom.

In this work, we discuss QED phenomena in a setup composed by quantum emitters and a Hofstadter-ladder waveguide.
Different from previous studies based on lattice environment with synthetic 
gauge fields~\cite{Burillo2020,WangX2020,Bernardis2021,Dong2021}, here we 
mainly focus on unconventional QED phenomena induced by the spin-orbit 
coupling. First, we assume that the emitter is of small atom form which 
frequency is resonant with the lower energy band. Due to 
spin-momentum locking, 
	the emitter chirally dissipates almost all its energy into one direction of 
	the 
	waveguide, which is different from the directional emission along the edge 
	states 
	of a 2D topologically non-trivial lattice~\cite{Mittal2014}. In our study 
	the chiral emission into the 1D Hofstadter-ladder waveguide stems from the 
	effective spin-orbit coupling, and 
	does not require any
	topological protection. Our proposal is possible to
	demonstrate chiral quantum optics, which has been extensively studied in 
	Refs.~\cite{Cala2019,Gheeraert2020,Wang2022arXiv,2021Dissimilar,Kannan2022On}.
Second, the emitter is considered as giant atom form, and couples to the 
waveguide at multiple 
sites~\cite{Kockum2014,Guo2017,Kockum2018,Kockum2019,Kannan2020,Zhao2020,DuL2021,ZhangY2021,WangC21,Dulei2022}.
Given that emitters' frequency is below two degenerate minima points induced by 
the spin-orbit coupling, there will be bound state in which the photonic energy 
will be trapped. The bound state induced by time-delay effects of 
	giant atom 
	has been investigated in Ref.~\cite{Guo2020}. In this work, we focus on 
	another giant atom
	effect, i.e., the quantum interference between 
	different coupling points. We find that, due to quantum interference and 
	unconventional spectrum of the	Hofstadter-ladder waveguide, the bound state 
	will be periodically modulated by the
	giant atom's size. The periodical length is tunable by 
	controlling the
	parameters of the Hofstadter ladder waveguide. Based on this mechanism, we 
	show 
that by tuning the interference as constructive/destructive, the dipole-dipole 
interaction between two giant emitters will be enhanced/suppressed.

\section{spectrum and spin-orbit coupling of Hofstadter-ladder waveguide}
\begin{figure}[tbph]
	\centering \includegraphics[width=8.4cm]{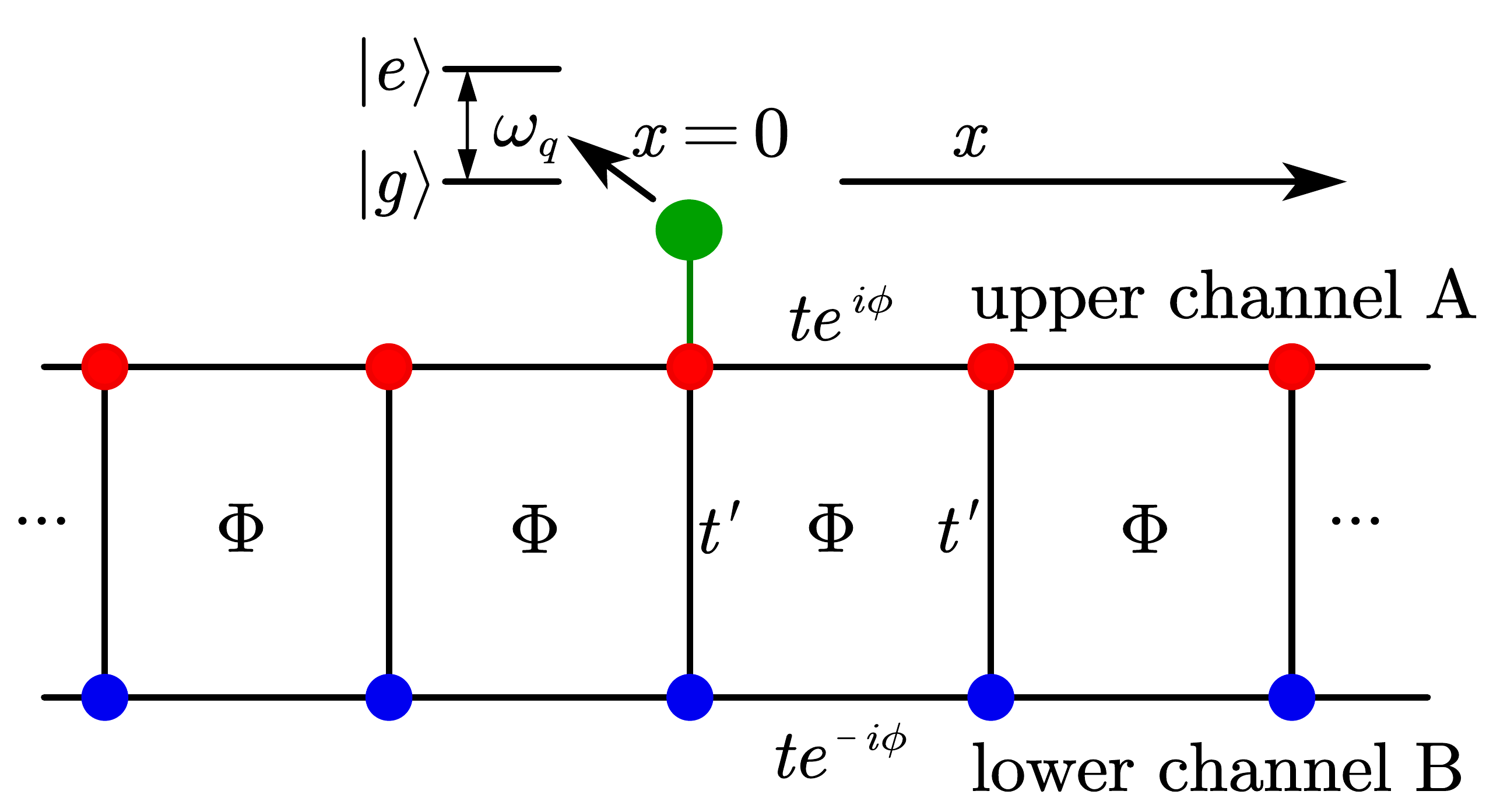}
	\caption{A two-level quantum emitter interacts with a waveguide in the form of the photonic analog of a Hofstadter-ladder model. The ladder rungs are hopped with strength $t'$. The nearest neighbor sites in two legs are coupled at rate $t$, together with a synthetic gauge phase $e^{i\phi}$ ($e^{-i\phi}$) for channel A (B). The effective magnetic field through each plaquette is $\Phi=2\phi$.}
	\label{fig1m}
\end{figure}

The model of the QED setup we study is depicted in Fig.~\ref{fig1m}, where a 
quantum emitter interacts with an artificial one-dimensional waveguide along 
the $x$ direction, which behaves as a photonic analog of the Hofstadter-ladder 
model.  The Hofstadter ladder can be viewed 
as the two-leg edge of the  
	Harper-Hofstadter model~\cite{Ozawa2019}, where a 
	synthetic gauge field 
	$\Phi=2\phi$ is applied through each plaquette (see 
	Fig.~\ref{fig1m}). Here we consider it working as a 1D artificial waveguide 
	which allows photons 
	traveling along it. In this situation, two legs in of the ladder waveguide 
	serve as 
	channel A and B of the 
	waveguide. 
For convenience, we set the length of one unit site as $d_0=1$. The ladder 
waveguide is composed by two legs, which can be viewed as two quantum channels 
for the emitter. Two sites in each rung are coupled with strength $t'$, which 
is set as 
$t'=1$. 
By adopting a Landau gauge along the $x$ direction, the phase connections only appear in each leg. Therefore, the hopping amplitude between two nearest neighbor sites is $te^{i\phi}$ ($te^{-i\phi}$) for channel A (B).
Consequently, by setting $\hbar=1$, the tight-binding Hamiltonian of the waveguide is~\cite{Guan2020} 
\begin{eqnarray}
	H_B&=&\sum_{x} \omega_0 (a_{x}^{\dagger}a_x+b_{x}^{\dagger}b_x) -\Big[t^{\prime}\sum_x{a_{x}^{\dagger}b_x} \notag \\
	&&+t\sum_x{\left(e^{i\phi}a_{x+1}^{\dagger}a_x+e^{-i\phi}b_{x+1}^{\dagger}b_x \right) }+\mathrm{H}.\mathrm{c.}\Big],
	\label{Hreal}
\end{eqnarray}
where $a_x, b_x$ ($a_{x}^{\dagger}, b_{x}^{\dagger}$) are the annihilation (creation) operators of the sites $a,b$ at position $x$, and $\omega_0$ is the identical frequency of those bosonic modes. In the following we work in the rotating frame of the constant part $\sum_{x} \omega_0 (a_{x}^{\dagger}a_x+b_{x}^{\dagger}b_x)$.

Under the periodic boundary condition and in the momentum space with 
\begin{gather}
a_{k}^{\dagger}=\frac{1}{\sqrt{N}}\sum_x{e^{ikx}a_{x}^{\dagger}}, \quad   b_{k}^{\dagger}=\frac{1}{\sqrt{N}}\sum_x{e^{ikx}b_{x}^{\dagger}}, \notag \\
 k=\frac{2\pi}{N}n,  \quad n\in (-N/2,N/2],
\end{gather}
we can diagonalize the waveguide Hamiltonian as 
\begin{eqnarray}
H_B&=&-2t\left[ \begin{matrix}
	a_{k}^{\dagger}	\quad	b_{k}^{\dagger} \notag \\
\end{matrix} \right] \mathcal{H}_B 
 \left[ \begin{array}{c}
	a_k\\
	b_k\\
\end{array} \right], \\
\mathcal{H}_B&=&\left[ \begin{matrix}
	g\left( k \right) +f\left( k \right)&		\eta\\
	\eta&		g\left( k \right) -f\left( k \right)\\
\end{matrix} \right] \notag  \\
&=&g\left( k \right)I+f\left( k \right)\sigma_z+\eta\sigma_x,
\label{H_B}
\end{eqnarray}
where $\eta=t^{\prime}/2t$, $g\left( k \right)$ and $f\left( k \right)$ are respectively expressed as 
\begin{equation}
g\left( k \right)=\cos  \phi  \cos  k , \quad  f\left( k \right)=\sin \phi  \sin  k .
\end{equation}

As shown in Eq.~(\ref{H_B}), the Hamiltonian $H_B$ is expressed in the Pauli operators, indicating that the upper-lower leg degree of freedom
behaves as an effective spin.
Due to the synthetic gauge field, $H_B$ contains the effective spin-orbit 
coupling term $(\sin \phi \sin k) \sigma_z$, which will lead to spin-momentum 
locking~\cite{Hugel2014}. Note that in condensed matter physics the 
concept 
	of ``spin" is extensively used for models consisting of``A" and ``B" 
	sublattices. 
	Similarly, the spin-orbit coupling is a generalized concept from atomic 
	physics, which describes a two-component internal freedom coupling to the 
	momentum of a particle~\cite{Ozawa2019}. For example, in 
	Refs.~\cite{Sala2015,Hafezi2011}, 
	the spin-orbit coupling and spin Hall insulators for photons have been 
	successfully demonstrated in experiments.

\begin{figure}[tbph]
 	\centering \includegraphics[width=8.4cm]{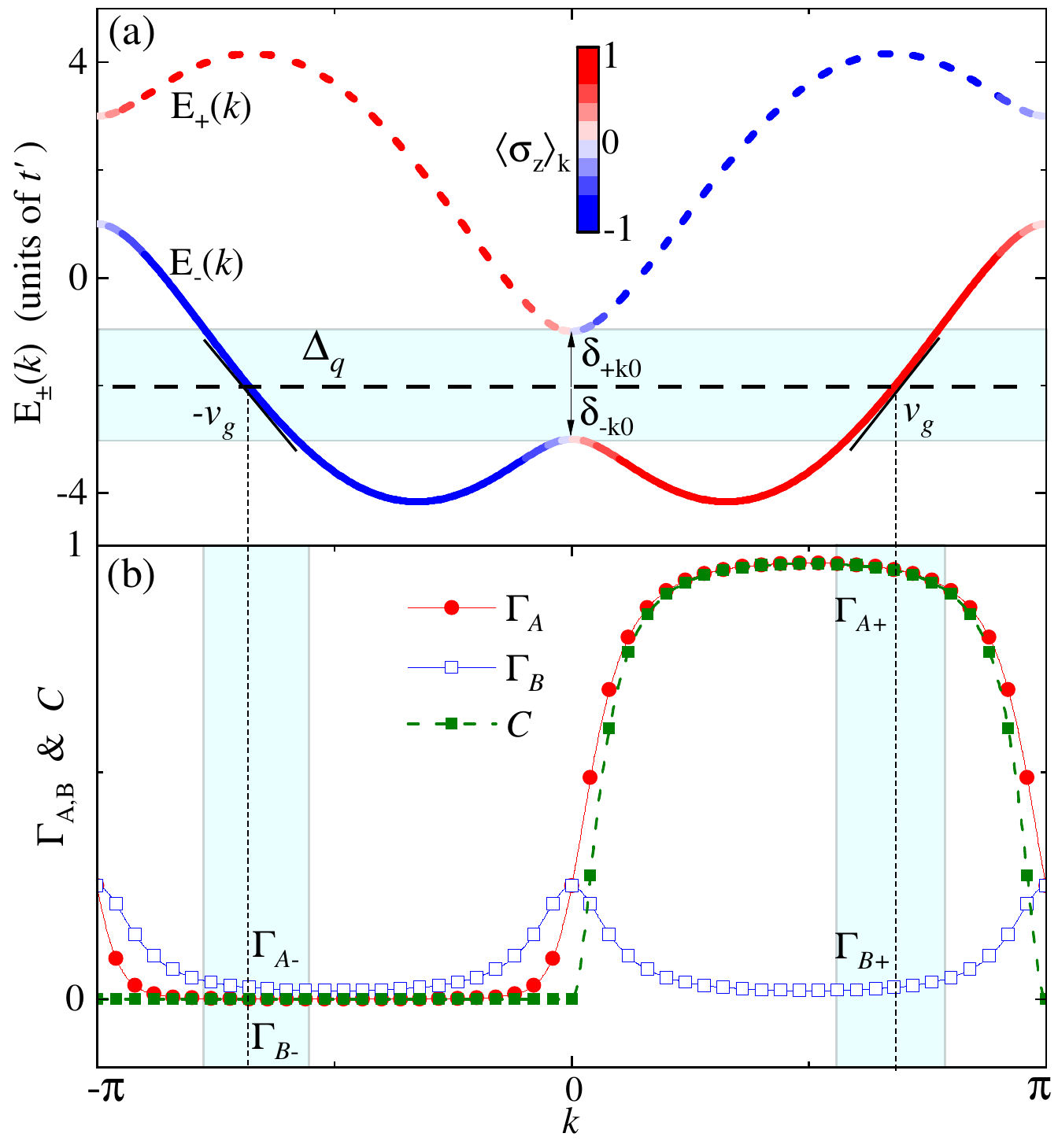}
 	\caption{(a) The dispersion relations for two energy bands of a 
 	Hofstadter-ladder waveguide. The effective spin $\langle \sigma 
 	_z\rangle_k$ describing the population difference between channel A and B, 
 	is mapped with colors. When discussing the chiral emission, the emitter's 
 	frequency is assumed in the cyan area, around which the group velocity is 
 	$v_g$. The detuning to the lower (upper) band edge is denoted by 
 	$\delta_{-k0}$ ($\delta_{+k0}$). (b) The analytical decay rates 
 	$\Gamma_{A(B)}$ [in the unit $g^2/(2v_g)$, see Eq.~(\ref{GammAB})] into 
 	channel A (B) and the chiral factor change with $k$. The cyan area 
 	corresponds to the Markovian decay regime where both band edges and the 
 	upper energy band do not take apparent effects. The cross points with the 
 	dashed vertical lines correspond to the decay rates of the emitter with 
 	frequency in (a). Parameters of the whole system: $t'=1$, $t=2$ and 
 	$\phi=\pi/3$.}
 	\label{fig2m}
\end{figure}

The energy spectrum can be derived by simply diagonalizing $\mathcal{H}_B$. Consequently, the energy bands and eigenmodes are derived as
\begin{gather}
	E_{\pm}\left( k \right) =-2t\left[ g\left( k \right) \mp\sqrt{f^2\left( k \right) +\eta^2} \right], \\      
	C_{k-}^{\dagger}=\left( \cos \frac{\theta _k}{2}
	a_{k}^{\dagger}	, \quad \sin \frac{\theta _k}{2}b_{k}^{\dagger}\right), \label{Ckm} \\
	C_{k+}^{\dagger}=\left( \sin \frac{\theta _k}{2}
	a_{k}^{\dagger}, \quad	-\cos \frac{\theta _k}{2}b_{k}^{\dagger}
	\right), \label{Ckp}
\end{gather}
where $\theta _k=\arctan[\eta/f(k)]$. Now we define the average spin as 
\begin{equation}
\langle \sigma _z\rangle_k = \langle a_{k}^{\dagger}a_{k}	\rangle-\langle b_{k}^{\dagger}b_{k}\rangle = \cos ^2\frac{\theta _k}{2}-\sin ^2\frac{\theta _k}{2}.
\end{equation}
Given that $\langle \sigma _z\rangle_k>0$ ($\langle \sigma _z\rangle_k<0$), the mode $k$ asymmetrically distributes on channel $A$ ($B$) with more probabilities. 
In Fig.~\ref{fig2m}(a), we plot $E_{\pm}\left( k \right)$ and $\langle \sigma _z\rangle_k$ of two bands versus $k$. It is found that,
due to spin-momentum locking, the chiral current $\langle \sigma _z\rangle_k$ of the lower band is opposite to the upper band for certain momentum $k$. 

Moreover, the spin-orbit coupling significantly modifies the dispersion relation of the waveguide.
For a free particle without spin-orbit coupling, the energy minimum point is usually with zero momentum $k=0$ (or $k=\pm\pi$).  When spin-orbit coupling appears, the spin-up and spin-down
modes minimize their energies by carrying non-zero opposite
momentum $k$~\cite{Galitski2013}. As depicted in Fig.~\ref{fig2m}, the photonic dispersion relation of the ladder waveguide becomes spin-dependent, and the energy minima are degenerate at two points with non-zero momentum $k\neq0$. Moreover, there is Kramers degeneracy for a pair of modes $\pm k$ due to spin-momentum interaction, and the field distribution is mostly localized in channel A (B) for $k>0$ ($k<0$)~\cite{Hugel2014}. Those properties allow us to realize unconventional phenomena of quantum optics, which will be addressed in the following discussions. 

\section{chiral spontaneous emission}
We first consider that the two-energy-level emitter is of small atom form, i.e., couples to the Hofstadter-ladder waveguide at one site $x=0$ (see Fig.~\ref{fig1m}). Its frequency lies resonantly within the lower energy band. In the rotating frame of bosonic frequency $\omega_0$, the system Hamiltonian is written as 
\begin{gather}
H_{s}=H_0+H_{\mathrm{int}}, \label{numH} \\
H_0=\frac{1}{2}\Delta_q\sigma _z+ H_B,  \quad 
H_{\mathrm{int}}=g\left( \sigma _-a_{0}^{\dagger}+\sigma _+a_0 \right),
\end{gather}
where $\Delta_q=\omega_q-\omega_0$ with $\omega_q$ being the emitter's transition frequency, $\sigma _z=|e\rangle\langle e|-|g\rangle\langle g|$ and $\sigma _+=(\sigma _-)^{\dagger}=|e\rangle\langle g|$, with $|e\rangle$ ($|g\rangle$) being the exited (ground) state of the emitter. Applying inverse Fourier transform, one obtains  $a^{\dagger}_0=\sum_k a_{k}^{\dagger}/\sqrt{N}$. 
According to Eqs.~(\ref{Ckm}, \ref{Ckp}), $a_{k}^{\dagger}$ can be decomposed as the superposition of $C_{k\pm}^{\dagger}$. Finally, the interaction Hamiltonian is written as 
\begin{eqnarray}
H_{\mathrm{int}}\!=\!\!\frac{g}{\sqrt{N}}
 \sum_k{\!\sigma _-\!\!\left( \!\cos \frac{\theta _k}{2}C_{k-}^{\dagger}\!+\!\sin \frac{\theta _k}{2}C_{k+}^{\dagger} \! \right) \!+\!\mathrm{H}.\mathrm{c}.}
\end{eqnarray}

As shown in Fig.~\ref{fig2m}(a), $\omega_q$ is set in the cyan regime, and only the lower band $E_{-}(k)$ is resonant with the emitter. To avoid the non-Markovian effects led by the band tops~\cite{Calajo2016}, we require $\Delta_q$ far away from two band edges, i.e., $|\delta_{\pm k0}|\gg0$.
By dropping the off-resonant terms with upper band modes $C_{k+}^{\dagger}$, the interacting Hamiltonian is reduced as 
\begin{equation}
H_{\mathrm{int}}=\frac{g}{\sqrt{N}}\left(\sum_k{\cos \frac{\theta _k}{2}\sigma _-\,\,C_{k-}^{\dagger}+\mathrm{H}.\mathrm{c}.} \right).
\label{Hint2}
\end{equation}
After substituting Eq.~(\ref{Ckm}) into Eq.~(\ref{Hint2}), we can divide $H_{\mathrm{int}}$ into two parts which describe interactions with channel A and B respectively:
\begin{gather}
H_{\mathrm{int}}=H_{\mathrm{int},A}+H_{\mathrm{int},B}, \label{HintABs} \\
H_{\mathrm{int},A}=
\frac{g}{\sqrt{N}}\left(\sum_k{\cos^2 \frac{\theta _k}{2}\sigma_-a_{k}^{\dagger}  +\mathrm{H}.\mathrm{c}.} \right), \label{HA}\\
H_{\mathrm{int},B}=
\frac{g}{\sqrt{N}}\left(\sum_k{\cos\frac{\theta _k}{2}\sin\frac{\theta _k}{2}\sigma_-b_{k}^{\dagger}  +\mathrm{H}.\mathrm{c}.} \right).  \label{HB}
\end{gather}
From Eq.~(\ref{HintABs}) and as depicted in Fig.~\ref{fig2m}(a), we find four dissipation terms by assuming the resonant position at $\Delta_q=E_{-}(\pm k_r)$, i.e., the left/right direction of channel A (B). After applying the unitary transformation $U_0(t)=\exp(-iH_0t)$, the interaction operator with channel A becomes 
\begin{equation}
\sum_k{( \sigma _-a_{k}^{\dagger})}\rightarrow \frac{N}{2\pi}\int_{-\pi}^{\pi}{\left( \sigma _-a_{k}^{\dagger}e^{i\Delta _kt} \right) dk}, \label{integral}
\end{equation}
where $\Delta _k=E_-\left( k \right) -\Delta_q$. Similar to Eq.~(\ref{integral}), the interaction operator with channel B can also be written in an integral form. We consider the spontaneous decay process with an excitation initially localized in the emitter. In the single-excitation subspace, the state of the whole system is expressed as $|\psi(t)\rangle=\sum_{k} [c_{ka}(t)|g,1_{ka}\rangle+c_{kb}(t)|g,1_{kb}\rangle]+c_{e}(t)|e,0\rangle$, and
the evolution of the whole system governed by $H_{\mathrm{int}}$ is derived from the following differential equations
\begin{gather}
	\dot{c}_{e}(t)=-i\sum_k  \frac{ge^{-i\Delta_k t}}{\sqrt{N}} \left[\cos^2 \frac{\theta _k}{2}c_{ka}(t)+\frac{\sin \theta _{k}}{2}c_{kb}(t)\right], \label{cet}  \\
	\dot{c}_{ka}(t)=-i\frac{g^*}{\sqrt{N}}e^{i\Delta_k t}\cos^2 \frac{\theta _k}{2}c_{e}(t), \label{cakt} \\
	\dot{c}_{kb}(t)=-i\frac{g^*}{\sqrt{N}}e^{i\Delta_k t}\frac{\sin \theta _{k}}{2}c_{e}(t).
	\label{cbkt}
\end{gather}
By substituting the internal form of Eqs.~(\ref{cakt}, \ref{cbkt}) into Eq.~(\ref{cet}), the evolution of $c_e(t)$ is derived as
\begin{widetext}
	\begin{equation}
		\dot{c}_e\left( t \right) =- \frac{g^2}{2\pi}\sum_{\pm}\left[\left( \frac{\cos \theta _{\pm kr}+1}{2} \right) ^2+
		\left( \frac{\sin \theta _{\pm kr}}{2} \right) ^2
		\right]\left|\int_0^{\pm \pi}{dk}\int_0^t{\left[ c_e\left( t^{\prime} \right) e^{-i\Delta _k\left( t-t^{\prime} \right)} \right] dt^{\prime}}\right|.
		\label{GammAB}
	\end{equation}
\end{widetext}

As depicted in Fig.~\ref{fig2m}(a), we approximate the dispersion relation 
around $\pm k_r$ to be linear, i.e., 
\begin{eqnarray}
v_g&=&\frac{dE_-\left( k \right)}{dk}\Big|_{k_r}  \notag \\
&=&
-2t\sin \left( k_r \right) \left(-\cos \phi  +\frac{\sin ^2 \phi  \cos k_r}{\sqrt{f^2\left( k_r \right) +\eta ^2}} \right) , \label{vg}
\end{eqnarray}
where $v_g$ is the group velocity at $k_r$ ($k_r>0$). By setting $\delta k=k- 
k_r$, the detuning is written as $\Delta _k\simeq v_g\delta k$. In the 
Born-Markovian regime, the decay rate is required to be much smaller than the 
band width $\{\delta_{+k0},\delta_{-k0}\}$, and we can extend the integral 
bound $\pm\pi$ to be infinite. Consequently, Eq.~(\ref{GammAB}) is reduced as 
\begin{equation}
\dot{c}_e\left( t \right) =-\sum_{i=A,B} \left(\Gamma_{i+}+\Gamma_{i-}\right) c_e\left( t \right),
\label{ex_decay}
\end{equation}
where $\Gamma_{i\pm}$ correspond to the emission rates into the right/left direction of channel $i$, which are derived as 
\begin{gather}
\Gamma_{A\pm}=\frac{g^2}{2v_g}\left( \frac{\cos \theta _{\pm kr}+1}{2} \right) ^2,  \label{Gamma_AB1} \\
\Gamma_{B\pm}=\frac{g^2}{2v_g}\left( \frac{\sin \theta _{\pm kr}}{2} \right) ^2,
\label{Gamma_AB}
\end{gather}
which show that $\Gamma_{A(B)\pm}$ are determined by $\theta _{\pm kr}$. In 
this part we only focus on the Markovian decay regime (cyan area in 
Fig.~\ref{fig2m}) where both band edges 
and the upper energy band do not take apparent effects. We plot $\Gamma_{A,B}$ 
in Fig.~\ref{fig2m}(b) [in units of 
$g^2/(2v_g)$], and find that the 
emission to channel A (B) is spatially asymmetric (symmetric), i.e.,
$$\Gamma_{A+}>\Gamma_{A-}, \qquad \Gamma_{B+}= \Gamma_{B-}.$$

Specially, under the following condition
\begin{equation}
\left(\frac{\cos \theta _{ kr}+1}{2} \right) ^2\gg \left\{ \left(\frac{\cos \theta _{-kr}+1}{2} \right) ^2, \left( \frac{\sin \theta _{\pm kr}}{2} \right) ^2   \right\},
\end{equation}
the emission field mostly distributes on the right side of channel A. 
Therefore, the spontaneous emission field will chirally propagate along the 
Hofstadter-ladder waveguide. As discussed in 
Sec.~II, the chirality is led by the effective spin-orbit coupling mechanism. 
\begin{figure}[tbph]
	\centering \includegraphics[width=8.8cm]{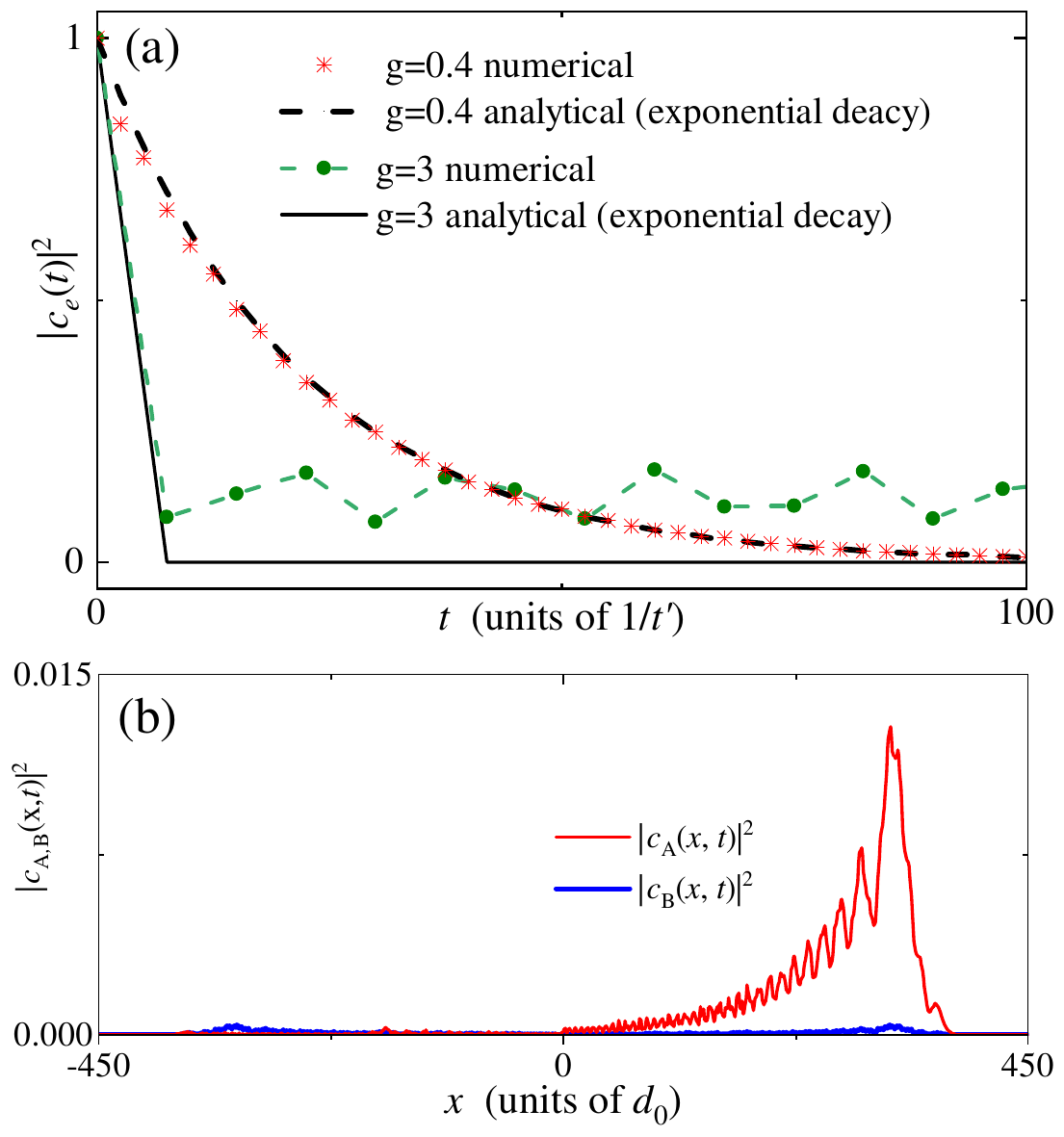}
	\caption{(a) The spontaneous decay of the emitter by setting $g=0.4$ and 
	$g=3$, respectively. The curves with (without) symbols correspond to 
	numerical calculations (Markovian exponential decay). (b) Given that 
	$g=0.4$, the field distributions along channel A and B at $t=100$. The 
	parameters are adopted the same with those in Fig.~\ref{fig2m}.}
	\label{fig3m}
\end{figure}

We assume that the coupling position is at $x=0$, and therefore, the field intensities on the right (left) side of channel A and B are defined as 
\begin{equation}
\Phi_{A(B)\pm}=\sum_{x=0}^{\pm N/2}|c_{A(B),x}|^2,
\label{phi_AB}
\end{equation}
where $c_{A(B),x}$ is the field amplitude of site $a(b)$ of the rung at $x$ of the ladder.
For example, if the right side of channel A is the desired direction, both the 
dissipation into channel B and into the left hand side of channel A will lead 
to photonic leakage. Consequently, the chiral factor $\mathcal{C}$ is defined 
as~\cite{Lodahl2017} 
\begin{equation}
\mathcal{C}=\frac{\Phi_{A+}}{\sum_{i=A,B} \left(\Phi_{i+}+\Phi_{i+}\right)}=\frac{\Gamma_{A+}}{\sum_{i=A,B} \left(\Gamma_{i+}+\Gamma_{i-}\right)}.
\end{equation}
By adopting expressions of $\Gamma_{A(B)\pm}$ in Eqs.~(\ref{Gamma_AB1},\ref{Gamma_AB}), the analytical chiral factor is derived as 
\begin{equation}
\mathcal{C}=\frac{\left( f\left( k_r \right) +|f\left(  k_r \right) | \right) ^2}{\left( f\left( k_r \right) +|f\left(  k_r \right) | \right) ^2+2\eta ^2},
\label{chiral_f}
\end{equation}
where we employ the relation $\tan\theta_k=\eta/f(k)$.
In Fig.~\ref{fig2m}(b), we plot chiral factor $\mathcal{C}$ changing with resonant wave number $k_r$. Given that $-\pi<\phi<0$, $f\left( k_r \right)=\sin \phi  \sin  k_r<0 $ (Note that we have restricted $k_r$ as positive, i.e., $k_r>0$). In this case, $\mathcal{C}=0$, indicating that 
the field hardly dissipates into the desired channel. When $0<\phi<\pi$, the chiral factor is simplified as 
\begin{equation}
\mathcal{C}=\frac{2f ^2\left( k_r \right) }{2f ^2\left( k_r \right)+\eta ^2}=\frac{1}{1+\frac{\eta ^2}{2f ^2\left( k_r \right)}}.
\end{equation}
Therefore, under the condition
\begin{equation}
\frac{\eta ^2}{2f ^2\left( k_r \right)}\ll 1\longrightarrow \mathcal{C}\rightarrow 1,  \label{cond1}
\end{equation}
most excitation energy will dissipate into the right side of channel A. Those 
discussions indicate that both
the waveguide's parameters and the resonant position $k_r$ will directly determine the chiral factor. 

By adopting the parameters in Fig.~\ref{fig2m}, we obtain 
\begin{equation}
\eta ^2/[2f ^2\left( k_r \right)]\simeq 0.058 \rightarrow \mathcal{C}\simeq 0.944,
\label{chi}
\end{equation} 
which is a high chiral factor to realize cascaded quantum networks with 
multiple nodes.
To verify our above analysis, in the following we numerically simulate the 
system's evolution by adopting the system's Hamiltonian in Eq.~(\ref{numH}). 
Note that the ladder's Hamiltonian is expressed in real space [see 
Eq.~(\ref{Hreal})], and the ladder length is set as $N=1000$, which is long 
enough to avoid field reflection by the bounds. Figure~\ref{fig3m}(a) shows the 
evolution of the emitter with $g=0.4$ and $g=3$, respectively. Given that 
$g=0.4$, the analytical decay rate is calculated as 
$\sum_{i=A,B}\left(\Gamma_{i+}+\Gamma_{i-}\right)\simeq 0.0014$ [according to 
Eqs.~(\ref{Gamma_AB1}, \ref{Gamma_AB})], which is much smaller than the band 
width $\delta_{\pm k0}$, and the Markovian approximation is valid. The 
emitter's evolution $|c_e(t)|^2$ is shown in Fig.~\ref{fig3m}(a), which decays 
with time and matches well with the 
analytical exponential form. In 
Fig.~\ref{fig3m}(b), we plot the field distributions at $t=100$ for both channel A and B, where the photonic field mostly distributes on the right side of channel A. 
By adopting Eq.~(\ref{phi_AB}) the numerical chirality is about $ 
\mathcal{C}\simeq 0.943$, which is very close to the analytical result derived 
in Eq.~(\ref{chi}). 

Note that all the above results for the chiral decay are derived within the Born-Markovian approximation. When the emitter-waveguide interaction strength is comparable to the band width $\{\delta_{+k0},\delta_{-k0}\}$, both modes around $k=0$ and two energy minima points $k_{\pm\text{min}}$ (see Fig.~\ref{fig2m} and Fig.~\ref{fig5m}) with zero group velocity will prevent the emitter from decaying. Partial excitation energy will be trapped around the coupling position in the form of bound states~\cite{Calajo2016}. By increasing the emitter-waveguide interaction beyond the Markovian regime with $g=3$, we plot the emitter's evolution in Fig.~\ref{fig3m}(a), which is no longer of exponential decaying form. After dissipating partial energy into the waveguide, the rest is trapped within the emitter. Therefore, to work as a well-performed cascaded quantum system, the emitter in each node should couple to the waveguide within the Markovian regime. Next, we will discuss the behavior of bound states of both small and giant emitters due to band edge effects.

As discussed in experimental work in 
Refs.~\cite{Underwood2012,Houck2012n}, the bosonic modes 
	in the Hofstadter-ladder waveguide can be made by 
	cavities or LC 
	resonators, which will experience decoherence led by 
	the noisy environment. The decay of each site is assumed in the
	Lindblad form, i.e.,
	\begin{equation}
		\mathcal{L}\rho=\kappa\sum_{O=a,b}\sum_{x}	O_x\rho 
		O_x^{\dag
		}-\frac{1}{2}(O_x^{\dag }O_x\rho-\rho O_x^{\dag 
		}O_x),
	\end{equation}
	where $\kappa$ is the photonic decay rate of each cite. 
	By setting $\kappa=0$ 
	and $0.01$ respectively, we plot the corresponding field 
	distributions along the waveguide 
	at $t=100$. When the photonic wavepacket propagates 
	along the dissipative waveguide, it will decay the 
	energy into the environment. We find
	that the amplitude of the photonic field becomes much 
	lower than the non-dissipative case. However, the field distribution is 
	still chiral, which is
	not affected by this local coherence.

\begin{figure}[tbph]
	\centering \includegraphics[width=8.8cm]{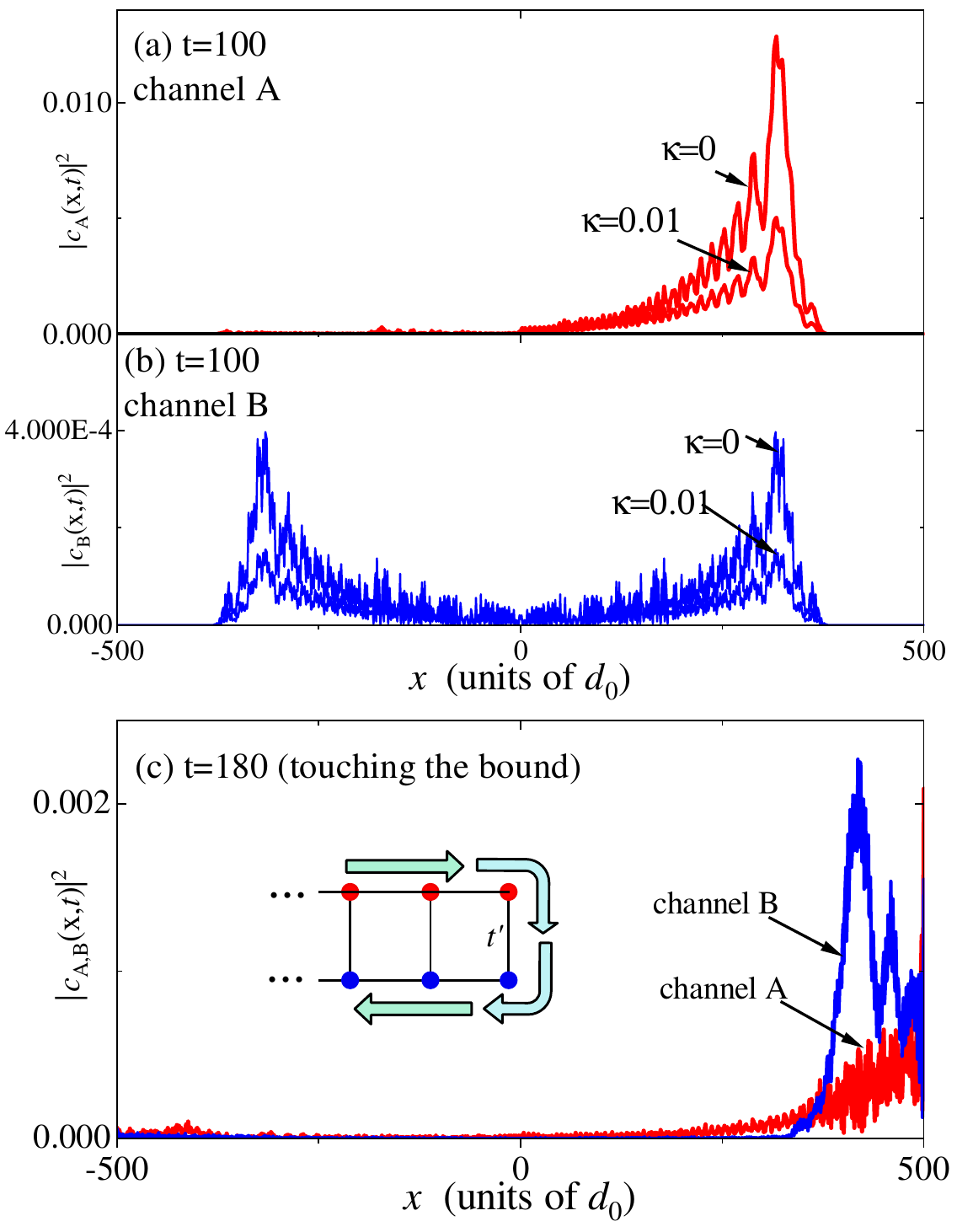}
	\caption{(a, b) By setting the photonic decay rate for 
	each site as $\kappa=0$ 
		and $\kappa=0.01$, the field distributions along 
		channel A and B at $t=100$. (c) By setting 
		$\kappa=0.01$ the field distributions after being 
		reflected by the hard-wall boundary of the 
		waveguide. The energy flow direction is shown in 
		the inset. Other parameters are adopted the same 
		with those in Fig.~\ref{fig3m}.}
	\label{fig4m}
\end{figure}

Additionally, in experiments the 
	waveguide's length cannot be infinite, and therefore, 
	the chiral field will be 
	reflected by the hard-wall boundary of the waveguide 
	after a long-time propagation. In Fig.~\ref{fig4m}(c), we 
	plot the field 
	distribution at $t=180$, when the wavepacket already 
	touches the waveguide's boundary.
	The energy flow direction is shown in the insets. We find that, due to 
	the hopping rates $t'$ between two channels at the 
	boundary, most of the energy will be reflected into 
	channel B. Since the photon decay of each rate is also 
	considered, the amplitude of the reflected wavepacket is 
	much lower than that in Fig.~\ref{fig4m}(a).

\section{periodical interference behavior modulated by giant emitter's size}
\subsection{bound state of a single giant emitter}

\begin{figure}[tbph]
	\centering \includegraphics[width=8.8cm]{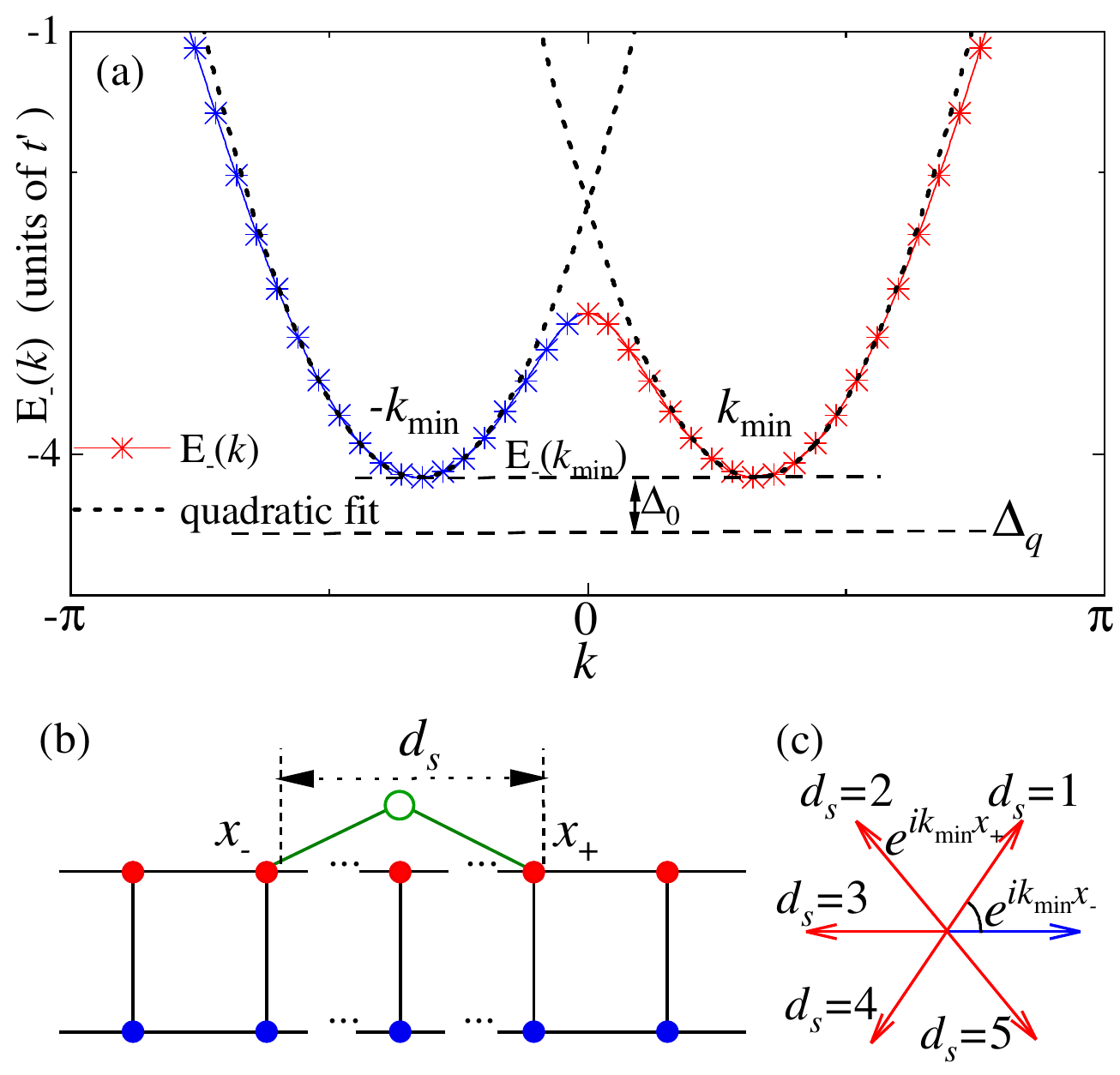}
	\caption{(a) Around two energy minima $\pm k_{\text{min}}$, the dispersion relation of the lower band can be fit with the quadratic relation [dashed curves, see Eq.~(\ref{Ek_q})]. To observe bound states, the emitter's frequency is set below the band edge, i.e., $\Delta_q<E_{k_{\text{min}}}$. (b) A giant emitter interacts with the waveguide at two coupling points $x_{\pm}$. The emitter's size is $d_s=x_{+}-x_{-}$. (c) Given that $k_{\text{min}}\simeq \pi/3$, the coupling phase relation between two coupling points changes with $d_s$.}
	\label{fig5m}
\end{figure}
Besides emitting and absorbing real photons, different emitters can also be coherently mediated via exchanging virtual photons in the waveguide, which requires the emitter's frequency to be outside the spectrum of the waveguide bath~\cite{Douglas2016,Bello2019,Wang2021}. Here we restrict the frequency detuning $\Delta_q$ below the lower bound of $E_{-}(k)$ [see Fig.~\ref{fig5m}(a)].
Additionally, the emitter is assumed to be of giant atom 
form~\cite{Kockum2014,Guo2017,Kockum2018,Kockum2019,Kannan2020,Zhao2020,DuL2021,ZhangY2021,WangC21,Dulei2022},
 which couples to the waveguide at two points $x_{\pm}$ of channel A (or B), as 
depicted in Fig.~\ref{fig5m}(b).
The separation distance is denoted as $d_s=x_+-x_-$, which corresponds to the giant emitter's size. Similar to previous discussions, the interaction strengths with channel A and B [see Eqs.~(\ref{HA},\ref{HB})] are written as 
\begin{eqnarray}
g_{ka}=\frac{g}{\sqrt{N}}\cos^2 \frac{\theta _k}{2}(e^{-ikx_-}+e^{-ikx_+}), 
\label{gkacos} \\
g_{kb}=\frac{g}{\sqrt{N}}\cos\frac{\theta _k}{2}\sin\frac{\theta _k}{2} 
(e^{-ikx_-}+e^{-ikx_+}),
\label{gkasin} 
\end{eqnarray}
where $g$ is the interaction strength with a single point. 
Note that under the condition $d_s=0$, two coupling positions coincide at the same site, and the giant atom degrades as a small atom.

Similar to Eqs.~(\ref{cet}-\ref{cbkt}), we can 
obtain differential 
	equations for $c_e(t)$ and $c_{ka(b)}(t)$. Defining $
	e^{i\Delta _kt}\tilde{C}_{ka\left( b \right)}\left( t 
	\right) =c_{ka\left( 
		b \right)}\left( t \right)$, the evolution is 
		derived in Laplace space with 
	$c_e(t)\rightarrow
	\tilde{c}_e(s)$ and $C_{ka(b)}(t)\rightarrow
	\tilde{C}_{ka(b)}(s)$ 
	~\cite{Calajo2016,Bello2019,Wang2022}
	\begin{gather}
		s\tilde{c}_e\left( s \right) =-i\sum_k{\left[ 
		g_{ka}\tilde{C}_{ka}\left( s \right) 
		+g_{kb}\tilde{C}_{kb}\left( s \right) \right]}, 
		\label{cess}\\
		s\tilde{C}_{ki}\left( s \right) =-i\Delta 
		_k\tilde{C}_{ki}\left( s \right) 
		-ig_{ki}^{*}\tilde{c}_e(s), \quad i=a,b.
	\end{gather}
	
	Consequently, $\tilde{C}_{ki}(s)$ is obtained as 
	\begin{equation}
		\tilde{C}_{ki}\left( s \right) 
		=\frac{-ig_{ki}^{*}c_e(s)}{s+i\Delta _k}, \quad i=a, 
		b.
		\label{C_ks}
	\end{equation}
	By substituting Eq.~(\ref{C_ks}) into Eq.~(\ref{cess}), 
	$\tilde{c}_e(s)$ is 
	derived as 
	\begin{gather}
		\tilde{c}_e\left( s \right) 
		=\frac{1}{s+\Sigma_{e}(s)},\\
		\Sigma 
		_e(s)=\sum_k{\frac{|g_{ka}|^2+|g_{kb}|^2}{s+i\Delta 
		_k}}, \label{self_S}
	\end{gather}
	where $\Sigma_{e}(s)$ is the self-energy. The 
	time-dependent evolution is 
	recovered from the inverse Laplace 
	transform~\cite{Ramos2016}
	\begin{equation}
		c_e\left( t \right)=\frac{1}{2\pi 
			i}\lim_{E\rightarrow\infty}\int_{\epsilon-iE}^{\epsilon+iE}
		\tilde{c}_e\left( s \right)e^{st}ds, \quad 
		\epsilon>0.
	\end{equation}
	Given that the waveguide is long enough to avoid 
	reflection 
	effects, we can 
	write the self-energy in the integral form by replacing 
	$\Sigma_k$ with
	$N/(2\pi)\int dk$. Substituting the relations
	\begin{gather}
		\cos ^2\frac{\theta _k}{2}=\frac{\frac{f\left( k 
		\right)}{\sqrt{f^2\left( k 
					\right) +\eta ^2}}+1}{2}, \\
		\frac{\sin \theta 
		_k}{2}=\frac{\frac{\eta}{\sqrt{f^2\left( k \right) 
		+\eta 
					^2}}}{2},
	\end{gather}
	into 
	Eq.~(\ref{self_S}), the self-energy is expressed as
	\begin{equation}
		\Sigma 
		_e(s)=\frac{1}{4\pi}\int_{-\pi}^{\pi}{\frac{\left(\frac{f\left(
		 k 
					\right)}{\sqrt{f^2\left( k \right) +\eta 
						^2}}+1\right)|G_k|^2}{s+i\Delta 
				_k}dk},
		\label{selfE}
	\end{equation} 
where $G_k=g(e^{-ikx_-}+e^{-ikx_+})$.

Due to the effective spin-orbit coupling, the energy minimum point is split 
into two with non-zero momentum $\pm k_{\text{min}}\neq 0$. At the positions of 
two dips, the group velocities are zero, which can be derived from 
Eq.~(\ref{vg})
\begin{equation}
v_g=\frac{dE_-\left( k \right)}{dk}\Big|_{k=\pm k_{\text{min}}}=0.
\end{equation}
Therefore, their positions are derived as
\begin{equation}
\sin \left( k_{\min} \right) =\pm \sqrt{\sin ^2\phi -\eta ^2\cot ^2\phi }.
\label{min_d}
\end{equation}
At $\pm k_{\min}$, the second-order derivatives are non-zero, and we denote 
the curvature as 
\begin{equation}
\alpha=\frac{d^2E_-\left( k \right)}{dk^2}\Big|_{k=\pm k_{\text{min}}}.
\end{equation}
As depicted in Fig.~\ref{fig5m}(a), we employ the effective mass approximation, 
to fit the dispersion around the band edges with quadratic relations
\begin{equation}
E_-\left( k \right) =E_{\min}+\alpha \left( k\pm k_{\min} \right) ^2.
\label{Ek_q}
\end{equation}

By substituting Eq.~(\ref{Ek_q}) into Eq.~(\ref{selfE}), the self-energy is 
calculated as
\begin{eqnarray}
\Sigma _e(s)&\simeq&
\frac{1}{4\pi}\Bigg\{\int_{-\pi}^{0}{\frac{\left(\frac{f\left( k 
\right)}{\sqrt{f^2\left( k \right) +\eta ^2}}+1\right)|G_k|^2}{s+i\left[\Delta 
_0+\alpha \left( k+ k_{\min} \right) ^2 \right]}dk} \notag \\
&&+\int_{0}^{\pi}{\frac{\left(\frac{f\left( k 
\right)}{\sqrt{f^2\left( k \right) +\eta 
^2}}+1\right)|G_k|^2}{s+i\left[\Delta_0+\alpha \left( k- k_{\min} \right) ^2 
\right]}dk} \Bigg\},
\label{selfE2}
\end{eqnarray}
where $\Delta _0=E_{\min}-\Delta_q$ is the detuning to the band edge [see 
Fig.~\ref{fig5m}(a)]. We assume that $\Delta _0$ is small, and therefore, only 
the modes 
around $\pm k_{\min}$ contribute significantly to $\Sigma _e(s)$. Consequently, 
we can approximate 
\begin{equation}
G_{\pm k}\simeq G_{\pm k_{\text{min}}}, \quad  f\left( \pm k \right)\simeq f( 
\pm 
k_{\text{min}}),	
\end{equation}
and extend the integral bound of Eq.~(\ref{selfE2}) to be infinite. Finally, 
we obtain 
\begin{equation}
\Sigma _e(s)\simeq \frac{|G_{k_{\text{min}}}|^2}{2}\frac{1}{\sqrt{-\alpha 
\left( \Delta _0-is \right)}}, \label{self_fi}
\end{equation}
where we have employed the following relations 
\begin{equation}
|G_{k_{\text{min}}}|=|G_{-k_{\text{min}}}|, \quad f( k_{\min})=-f(-k_{\min}).
\end{equation}
\begin{figure}[tbph]
	\centering \includegraphics[width=8.8cm]{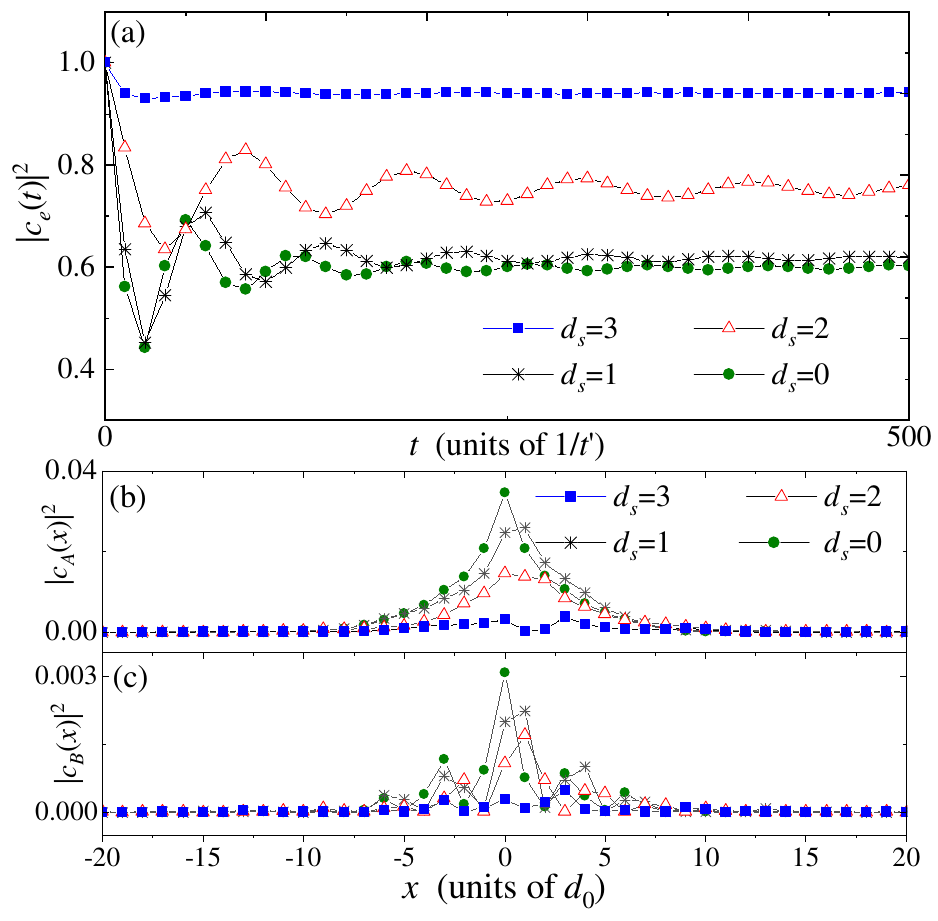}
	\caption{(a) The probabilities $|c_e(t)|^2$ of the giant emitter remaining 
	in its excited state change with $t$ for different $d_s$. In 
	the limit 
	$t\rightarrow \infty$, $|c_e(t)|^2$ reaches its steady value, which is 
	nonzero due to the band edge effect. The corresponding field distributions 
	of 
	the photonic bound states in channel A and B are shown in (b) and (c), 
	respectively. The emitter frequency is set below the lower band with 
	$\Delta_q=-4.2$. The coupling strength is $g=0.1$. The waveguide parameters 
	are the same 
	with those in Fig.~\ref{fig2m}.}
	\label{fig6m}
\end{figure}

In this work, the adopted parameters of the waveguide satisfy
\begin{equation}
\sin ^2\phi \gg \eta ^2 \cot ^2 \phi  \longrightarrow
k_{\text{min}}\simeq \phi= \pi/3.
\end{equation}
As shown in 
Fig.~\ref{fig5m}(c), the phase relation between two coupling points will rotate
counter-clockwise when increasing giant emitter's size ($d_s=0,1,2,3...$). 
Therefore, the relative phase between $x_{\pm}$ is very 
essential for the dynamics of the giant emitter. The interference between two 
points is maximum destructive (constructive) given that $d_s=3(2N+1)$ 
($d_s=6N$). When $\Delta_0$ is much 
stronger than $G_{k_{\text{min}}}$, most energy will be trapped in the 
emitter in the form of the bound 
state~\cite{Calajo2016,GonzlezTudela2017,Wang2022}. The trapped excitation 
probability is determined by the pure imaginary pole $s_0$ of the following 
transcendental 
equation~\cite{Wang2022}
\begin{equation}
s_0+\Sigma _e(s_0)\simeq 
s_0+\frac{|G_{k_{\text{min}}}|^2}{2}\frac{1}{\sqrt{-\alpha \left( \Delta 
_0-is_0 \right)}}=0, \label{root_s0}
\end{equation}
and the steady state population of the emitter is derived via the residue 
theorem~\cite{Calajo2016}
\begin{gather}
|c_e(t=\infty)|^2=|\text{Res}(s_{0})|^2, \\	\text{Res}(s_{0})=\frac{1}{1+\partial_s \Sigma_{e}(s)}\Big|_{s=s_{0}}.
\label{res}
\end{gather}

In Fig.~\ref{fig6m}(a), we plot dynamical evolutions of $|c_e(t)|^2$ for 
different giant emitter's sizes $d_s$. When $d_s=0$ (a small emitter), the 
effective interacting strength $|G_{k_{\text{min}}}|$ is strongest, which is 
enhanced by the constructive interference between two coupling legs. The steady 
state $|c_e(t)|^2$ reaches
lowest, and the emitter can effectively distribute its energy into the 
waveguide. That is, 
the photonic bound state is mostly localized in channel A due to the 
relation 
\begin{equation}
\cos^2 \frac{\theta _{k_{\text{min}}}}{2}\gg \cos\frac{\theta 
_{k_{\text{min}}}}{2}\sin\frac{\theta _{k_{\text{min}}}}{2},
\end{equation}
which can be seen clearly by comparing Fig.~\ref{fig6m}(b) and 
Fig.~\ref{fig6m}(c).

\begin{figure}[tbph]
		\centering \includegraphics[width=8.5cm]{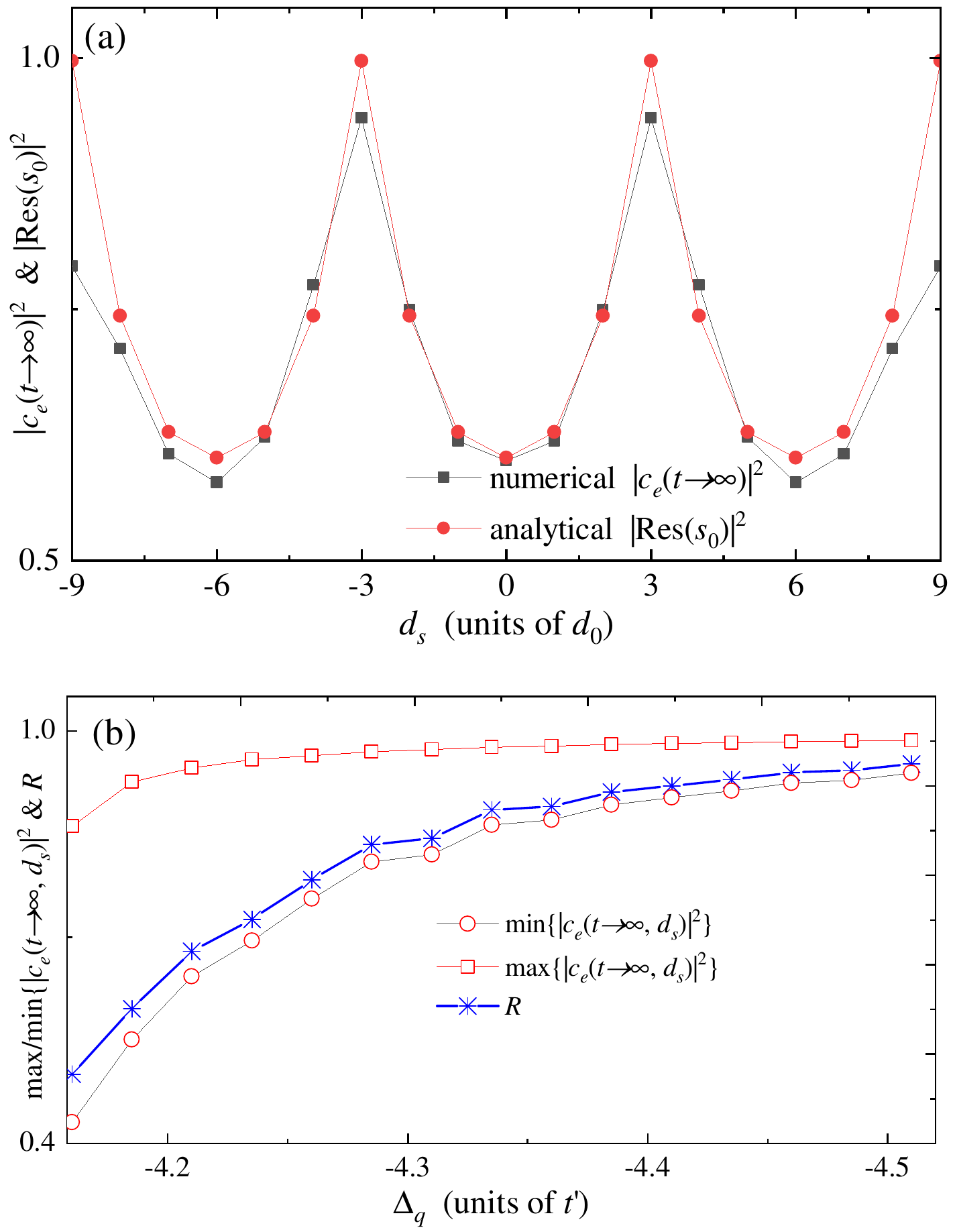}
		\caption{(a) The periodical behavior of the trapped excitation 
			$|c_e(t\rightarrow\infty)|^2$ changing with the size of giant 
			emitters. The 
			analytical results are calculated via the residue theorem in 
			Eq.~(55). (b) The maximum/minimum $|c_e(t\rightarrow\infty, d_s)|^2$ 
			and the contrast ratio $R$ changes with the emitter frequency 
			$\Delta_q$.}
		\label{fig7m}
	\end{figure}

When increasing the distance between two coupling points, the constructive 
interference will be reversed as destructive, with $|G_{k_{\text{min}}}|$ 
being significantly weaken. When $d_s=3$, the phase difference satisfies 
$k_{\text{min}}d_s\simeq \pi$, indicating that the interaction strength is 
approximately zero, i.e., $|G_{k_{\text{min}}}|\simeq 0$ [see 
Fig.~\ref{fig5m}(c)]. The giant emitter 
approximately 
decouples with the waveguide. Due to this decoupling mechanism, the excitation 
trapped in the emitter reaches its maximum, as shown in Fig.~\ref{fig6m}(a). 
Figure~\ref{fig6m}(b, c) show that the destructive interference will also 
suppress the bound state's amplitude significantly.
We plot the steady state population 
$|c_e(t\rightarrow\infty)|^2$ changing with $d_s$ in Fig.~\ref{fig7m}(a), which 
clearly presents a 
periodical interference pattern of $|c_e(t\rightarrow\infty)|^2$ modulated by 
the giant atom's size. Note that we only take 
$k_{\text{min}}\simeq \pi/3$ for example. Note that 
The periodical length is tunable by 
controlling the
parameters of the Hofstadter ladder waveguide.
By adopting another 
$k_{\text{min}}$ according Eq.~(\ref{min_d}), different spatial interference 
patterns can also be observed. The oscillating pattern in Fig.~\ref{fig7m}(a) 
is due to 
interference between different points in the giant 
	atom, and the peaks (dips) correspond to the positions where the 
	maximum constructive (destructive) interference happens. Therefore, we can 
	define the contrast ratio $R$ for the interference as
	\begin{equation}
		R=\frac{\min \{|c_e(t\rightarrow\infty,d_s)|^2\}}{\max 
		\{|c_e(t\rightarrow\infty,d_s)|^2\}}.
	\end{equation}
	For the parameters employed in Fig.~\ref{fig7m}(a), $\min 
	\{|c_e(t\rightarrow\infty,d_s)|^2\}$ ($\max 
	\{|c_e(t\rightarrow\infty,d_s)|^2\}$) is located at $d_s=0$ ($d_s=\pm3$),
	which corresponds to a small (giant) atom.
	A smaller $R$ indicates that the dynamical difference between small and 
	giant atoms is more apparent. In 
	Fig.~\ref{fig7m}(b), we plot maximum/minimum 
	$|c_e(t\rightarrow\infty,d_s)|^2$ and $R$ changing with the frequency 
	detuning $\Delta_q$. When shifting $\Delta_q$ away from the band edge, the 
	contributions of the band-edge modes to the bound states will be 
	significantly suppressed. Therefore, a larger 
	$\Delta_q$ 
	will reduce the interference effects, which corresponds to $R\rightarrow 1$ 
	[see Fig.~\ref{fig7m}(b)]. To observe 
	better interference effect in experiments, the emitter frequency can be  
	close to the edge of the lower energy band.
	
\subsection{dipole-dipole interactions}
\begin{figure}[tbph]
	\centering \includegraphics[width=8.8cm]{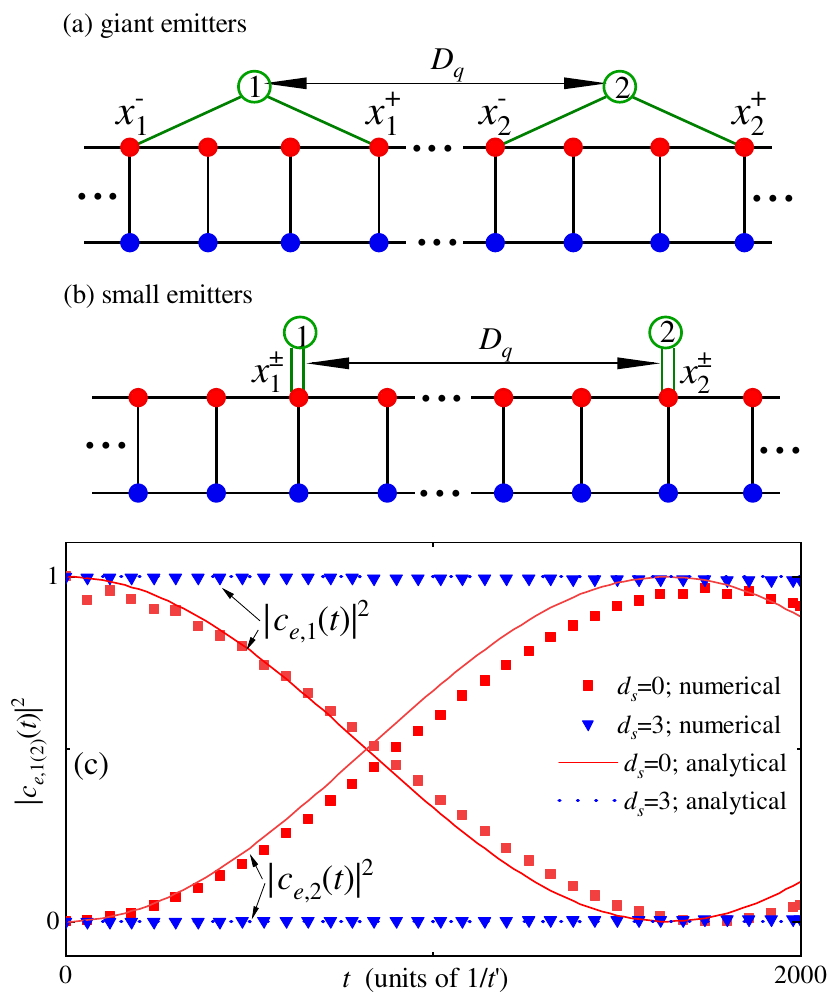}
	\caption{(a) Two giant and (b) two small emitters, which are separated with 
	distance $D_q$, interact with a common ladder waveguide. For giant emitter 
	$i$, the coupling points are located at different positions $x_{i}^{\pm}$. 
	For small emitter $i$, $x_{i}^{\pm}$ are coincided at the same site. (c) 
	The Rabi oscillations between two giant (small) emitters which are marked 
	with triangles (squares). The corresponding analytical results (curves 
	without symbols) are derived from Eq.~(\ref{J12}). The separation distances 
	between two emitters in (a, b) are set as $D_q=4$. To 
	avoid virtual photons in the waveguide being excited with high 
	probabilities, 
	we adopt a weak coupling strength $g=0.015$. The other parameters are the 
	same with those in Fig.~\ref{fig6m}.}
	\label{fig8m}
\end{figure}

By exchanging virtual photons in the waveguide, between emitters there are 
long-range dipole-dipole interactions which are determined by the overlap areas 
between their bound states~\cite{GonzlezTudela2015,Douglas2016,Bello2019}.  
As shown in Fig.~\ref{fig8m}(a, b), we now discuss the multiple emitters 
interacting with the same ladder waveguide. Due to the interference mechanism 
presented above, we focus on revealing the relation between giant emitter's 
size and the quantum dynamics where dipole-dipole interactions are involved.

Both two giant emitters are assumed to
couple with channel A, and the coupling topology is of separation 
form~\cite{Kockum2018}. Similar 
to the single emitter case, the interaction 
Hamiltonian is written as 
\begin{equation}
	H_{\mathrm{int},2}=\sum_{i=1,2}\sum_k\frac{G_{ik}}{\sqrt{N}}{\cos 
	\frac{\theta _k}{2}\sigma^-_{i}\,\,C_{k-}^{\dagger}+\mathrm{H}.\mathrm{c}.},
	\label{Hint2qo}
\end{equation}
where $G_{ik}=g(e^{-ikx^i_-}+e^{-ikx^i_+})$ is the coupling strength between 
emitter $i$ and the waveguide. We consider two emitters with identical 
frequency $\Delta_q$ which is also below the lower bound of $E_{-}(k)$. To 
proceed, we define the average distance between two emitters as
\begin{equation}
	D_q=\frac{x^2_++x^2_-}{2}-\frac{x^1_++x^1_-}{2}.
\end{equation}

The dipole-dipole interaction can be tediously 
derived via the standard
	resolvent-operator 
	techniques~\cite{GonzlezTudela2015,Douglas2016,Bello2019,Wang2021}.
	 However, given that the detuning $\Delta_0$ is large 
	and the waveguide is only virtually excited, $J_0$ 
	corresponds to the effective coupling strength mediated 
	by the waveguide's modes, which can be simply derived 
	via the effective Hamiltonian methods~\cite{James2007}. 
	In the rotating frame of free energies of both emitters and 
	waveguide, we first write $H_{\mathrm{int},2}$ in 
	Eq.~(\ref{Hint2qo}) in the time-dependent form 
	\begin{equation}
		H_{\mathrm{int},2}(t)=\sum_{i=1,2}\sum_k\frac{G_{ik}}{\sqrt{N}}{\cos
			\frac{\theta_k}{2}e^{i\Delta_{k}t}\sigma^-_{i}\,\,C_{k-}^{\dagger}+\mathrm{H}.\mathrm{c}.},
		\label{Hint2q}
	\end{equation}
	where $\Delta_{k}=E_{-}(k)-\Delta_{q}$. We first 
	calculate the coupling 
	rate $J_{12,k}$ mediated by one mode $k$. By employing the methods in 
	Ref.~[76], the one-mode-mediated effective Hamiltonian is    
	derived as
	\begin{eqnarray}
		H_{q-q,k}\!&=&\!\frac{G_{1k}G^*_{2k}}{\Delta_{k}}\frac{\cos^2
			\frac{\theta _k}{2}}{N}\times \notag \\
		&&\!\left( 
		\sigma^-_{1}C_{k-}^{\dagger}\sigma^+_{2}C_{k-}\!-\!\sigma^+_{2}C_{k-}\sigma^-_{1}C^{\dagger}_{k-}\right)\!+\!\mathrm{H}.\mathrm{c}.
	\end{eqnarray}
	Since the waveguide is only virtually excited and 
	approximately in its vacuum state, we can trace off the 
	freedom of mode $k$ by adopting the following 
	approximation 
	\begin{equation}
		\langle C_{k-}^{\dagger}C_{k-}\rangle\simeq 0, \quad 
		\langle C_{k-}C_{k-}^{\dagger}\rangle\simeq 1.
	\end{equation}
	Consequently, $H_{q-q,k}$ is simplified as 
	\begin{equation}
		H_{q-q,k}\simeq 
		-\frac{G_{1k}G^*_{2k}}{\Delta_{k}}\frac{\cos^2 
			\frac{\theta 
			_k}{2}}{N}\sigma^+_{2}\sigma^-_{1}+\mathrm{H.c.}
	\end{equation}
	Note that $H_{q-q,k}$ is mediated by one mode, and 
	the total 
	dipole-dipole interaction should take all the modes' 
	contribution into 
	account. Consequently, the total dipole-dipole interaction Hamiltonian 
	mediated by all the waveguide's modes is derived as 
	\begin{equation}
		H_{q-q}=\sum_kH_{q-q,k}=J_{12}\left( \sigma 
		_{1}^{-}\sigma _{2}^{+}+\mathrm{H}.\mathrm{c}. 
		\right).
	\end{equation}
	where $J_{12}$ is the total interaction strength which is 
	written as 
	\begin{eqnarray}
		J_{12}&=&-\sum_k 
		\frac{G_{1k}G^*_{2k}}{\Delta_{k}}\frac{\cos^2 
			\frac{\theta _k}{2}}{N} \notag \\
		&\simeq& 
		-\frac{1}{2\pi}\int_{-\pi}^{\pi}\frac{G_{1k}G^*_{2k}\cos^2
			\frac{\theta _k}{2}}{\Delta_{k}}dk.
	\end{eqnarray}
	Since the emitter frequency is below the edge of the 
	lower energy band 
	[see Fig.~\ref{fig5m}(a)], the dispersion relation can 
	be fit with the 
	quadratic relations in Eq.~(\ref{Ek_q}). Without loss of 
	generality, we set
	$(x^1_++x^1_-)/2=0$ since only relative distance
	matters. As depicted in Fig.~8(a), two emitters are separated with a 
	distance $D_q$. Given that 
	$D_q>d_s$, $J_{12}$ can be written as
	\begin{eqnarray}
		J_{12}&\simeq& 
		-\frac{1}{2\pi}\Bigg\{\int_{-\pi}^{0}\frac{|G_{1k}||G^*_{2k}|e^{ikD_q}\left(
		 \frac{\cos \theta _{k}+1}{2} \right)}{\Delta 
			_0+\alpha \left( k+ k_{\min} \right) ^2 }dk 
			\notag \\
		&+&\int_{0}^{\pi}\frac{|G_{1k}||G^*_{2k}|e^{ikD_q}\left(
		 \frac{\cos \theta _{k}+1}{2} \right)}{\Delta 
			_0+\alpha \left( k- k_{\min} \right) ^2 }dk 
			\Bigg\} \label{J120}
	\end{eqnarray}
	As shown in Eq.~(\ref{Gamma_AB1}) and depicted in 
	Fig.~\ref{fig2m}(b), the 
	following relation
	$$    |\frac{\cos \theta _{-k}+1}{2}|\ll|\frac{\cos 
	\theta _{k}+1}{2}|\simeq 1, \quad k>0$$ 
	is valid for the parameters employed in this work. 
	Consequently, we can neglect the first part ($k<0$) in
	Eq.~(\ref{J120}), and $J_{12}$ is derived as 
	\begin{eqnarray}
		J_{12}\simeq 
		\int_{0}^{\pi}\frac{|G_{1k}||G^*_{2k}|e^{ikD_q}}{\Delta
			_0+\alpha \left( k- k_{\min} \right) ^2 }dk,
	\end{eqnarray}
	which indicates that
	that dipole-dipole interactions are 
	determined by the overlap areas 
	between two bound states of emitters~\cite{GonzlezTudela2015}. 
	Similar to the process obtaining the self-energy in 
	Eqs.~(\ref{selfE2}, \ref{self_fi}), we derive $J_{12}$ 
	as 
	\begin{equation}
		J_{12}=\frac{|G_{1k\min}||G_{2k\min}^{*}|}{2\sqrt{\alpha
		 \Delta _0}}e^{-|\sqrt{\frac{\Delta 
		_0}{\alpha}}D_q|},
		\label{J12}
	\end{equation}
which shows that $J_{12}$ exponentially decays with emitters' 
separation distance. Similar results have been obtained in 
Refs.~\cite{GonzlezTudela2015,Douglas2016,Wang2021}. 

As shown in Fig.~\ref{fig7m}(b, c), 
the amplitude of the bound state will periodically change with $d_s$ due to 
the interference effect. When $d_s=3$, the photonic bound state 
approximately disappears. Consequently, the overlapping area is also nearly 
zero. By assuming the initial excitation being in emitter 
1, we numerically plot the Rabi oscillations between giant emitters ($d_s=3$) 
and small emitters ($d_s=0$) 
in Fig.~\ref{fig8m}(c), respectively.
We find that, due to the destructive interference, when $d_s=3$, two emitters 
hardly 
exchange energy, and decouple with each other ($J_{12}\simeq 0$). On the 
contrary, two emitters will exchange excitation rapidly at a large rate  
due to the constructive interference when $d_s$ is reduced to be zero. 

\section{Conclusion and outlooks}
In this work, we explore the unconventional quantum optics by considering  
a Hofstadter-ladder waveguide interacting with both small and giant emitters. 
Due to the effective spin-orbit coupling, both the waveguide's vacuum modes and 
spectrum show nontrivial properties. In the first part, we consider a small emitter 
which frequency is resonant with one energy band. Two legs of the ladder, which 
can be viewed as the freedoms in an effective spin locked with the momemtum 
freedom, provide two dissipating channels. The spontaneous emission is chiral 
with most photonic field decaying unidirectionally. Both numerical and 
analytical results show that the Hofstadter-ladder waveguide can work as a 
well-performed quantum bus of a chiral network.

In the second part, the emitters are assumed of giant atom form with 
frequencies below the lower energy band. In this scenario, only the modes 
around two energy minima points induced by the spin-orbit coupling contribute 
significantly to the system's dynamics, which will lead to emitter-waveguide 
bound states.
Since the energy minima modes carry non-zero momentum, the coupling strengths are periodically modulated by the giant emitter's size due to quantum interference. Specially the giant emitter will decouple with the waveguide when maximum destructive interference happens. This mechanism provides a novel approach to control long-range dipole-dipole interactions between emitters by modulating their sizes. 

In this work, the giant emitter is assumed to couple with the sites of the same 
channel. Other intriguing effects might be observed by considering the coupling 
points on different channels. Moreover, for multiple giant emitters, we just 
consider the coupling points arranged in 
the sepration form. In fact there are some other distinct 
topologies, which are nested and braided cases~\cite{Kockum2018}. Exploring the coupling topology effects might also bring novel quantum phenomena.
There are already plenty of work about realizing artificial Hofstadter ladders 
in various quantum systems such as cold atoms and 
circuit-QED~\cite{Tai2017,Gu2017,Guan2020,Weitenberg2021},  which can be 
potential platforms to demonstrate above unconventional QED phenomena.
We hope that our work will open the possibilities of exploring novel quantum 
effects in artificial spin-orbit-coupling environments.

\section{Acknowledgments}
The quantum dynamical simulations are based on open source code 
QuTiP~\cite{Johansson12qutip,Johansson13qutip}. 
X.W.~is supported by
the National Natural Science
Foundation of China (NSFC) (Grant No.~12174303 and No.~11804270), and China Postdoctoral Science Foundation No.~2018M631136.
H.R.L. is supported by the National Natural Science
Foundation of China (NSFC) (Grant No.11774284).

%

\end{document}